\documentstyle[12pt]{article}
\makeatletter

\ifcase \@ptsize
\or % mods for 11 pt
\or % mods for 12 pt
\fi
\advance\textheight by \topskip

\ifcase \@ptsize
\or % mods for 11 pt
\or % mods for 12 pt
\fi

\def\section{\@startsection {section}{1}{\z@}{-3.5ex plus -1ex minus
 -.2ex}{2.3ex plus .2ex}{\large\bf\centering}}
\def\subsection{\@startsection{subsection}{2}{\z@}{-3.25ex plus -1ex minus
 -.2ex}{1.5ex plus .2ex}{\sc}}
\gdef\@publabel{\hfil}
\gdef\@pubdate{\null}
\gdef\@pubnumber{\null}
\gdef\@author{\null}
\gdef\@title{\null}
\gdef\@abstract{\null}
\long\def\pubdate#1{\gdef\@pubdate{#1}}
\long\def\pubnumber#1{\gdef\@pubnumber{#1}}
\long\def\publabel#1{\gdef\@publabel{#1}}
\long\def\author#1{\gdef\@author{#1}}
\long\def\title#1{\gdef\@title{#1}}
\long\def\abstract#1{\gdef\@abstract{#1}}
\def\titlerelax{
\let\maketitle\relax
\let\settitleparameters\relax
\let\consolidatetitle\relax
\let\inittitlepage\relax
\let\finishtitlepage\relax
\let\titlepagecontents\relax
\let\multithanks\relax
\let\titlebaselines\relax
\let\@makepub\relax
\let\@maketitle\relax
\let\@makeauthor\relax
\let\@makeabstract\relax
\let\@maketitlenote\relax
\let\thanks\relax
\let\titlerelax\relax}
\def\titleclean
{\gdef\@titlenote{}
\gdef\@abstract{}
\gdef\@author{}
\gdef\@title{}
\gdef\@pubdate{}\gdef\@pubnumber{}\gdef\@publabel{}
\gdef\@dpublabel{}
}
\def\@makepub{\vbox to \z@{\hbox to \textwidth{\hfill
\@publabel \hfill
\llap{\parbox[t]{0.25\textwidth}{\raggedleft\@pubnumber}}}%
\vss}}
\def\@maketitle{\vskip 60pt \begin{center}
 {\LARGE \@title \par}
 \end{center}}
\def\@makeauthor{{\def\and{\smallskip {\normalsize \rm and\smallskip}}
\long\def\address##1{{\def\and{\\and\\}\medskip
				{\small \it \\##1\\}
}}
{\centering
 \vskip 3em
 \large \lineskip .75em
 \@author}
 \par}}
\def\@makedate{\vskip 1.5em
 {\raggedright \small \noindent\@pubdate \par}}
\def\@makeabstract{\vskip 1.5em
{\small
\begin{center}
{\bf ABSTRACT\vspace{-.5em}\vspace{0pt}}
\end{center}
\quotation \@abstract \endquotation}}
\def\maketitle{\titlepage
\let\footnotesize\small \setcounter{page}{0}
\@makepub
\vfil
\@maketitle
\@makeauthor
\vfil
\@makeabstract
\@thanks
\vfil
\@makedate
\if@restonecol\twocolumn \else \eject \fi
\titlerelax \titleclean
\setcounter{footnote}{0}
}

\def\@cite#1#2{\nolinebreak$^{[\scriptstyle #1\if@tempswa , #2\fi]}$}
\def\@citex[#1]#2{\if@filesw\immediate\write\@auxout{\string\citation{#2}}\fi
  \def\@citea{}\@cite{\@for\@citeb:=#2\do
    {\@citea\def\@citea{,\penalty\@m}\@ifundefined
       {b@\@citeb}{{\bf ?}\@warning
       {Citation `\@citeb' on page \thepage \space undefined}}%
{\csname b@\@citeb\endcsname}}}{#1}}

\@definecounter{footnote}

\newcommand{\hsp}{\hspace{0.08in}}

\begin{document}
\newcommand{\A}{{\cal A}}
\bibliographystyle{npb}
\pubdate{\today}
\pubnumber{DTP-92-33}
\title{The full set of $c_n$-invariant factorized $S$-matrices}
\author{N. J. MacKay\footnote{Supported by a U.K. Science and Engineering
Research
Council studentship. Address from 1st August: RIMS, Kyoto Univ., Kyoto 606,
Japan.}\address{
        Department of Mathematical Sciences, \linebreak
        University of Durham, \linebreak
        Durham, DH1 3LE, England}}

\abstract{
We use the method of the tensor product graph to construct rational (Yangian
invariant) solutions of the Yang-Baxter equation in fundamental representations
of $c_n$ and thence the full set of $c_n$-invariant factorized $S$-matrices.
}

\maketitle
\baselineskip 20pt
\parskip 0.2in
\parindent 12pt

Integrable quantum field theories in 1+1 dimensions are expected to have exact
$S$-matrices in which particle number and the set of asymptotic momenta are
conserved, and in which multiparticle interactions factorize into products of
two-particle interactions. The condition that this factorization be consistent
is then the Yang-Baxter equation (YBE), so that in theories with a global Lie
group invariance, such as the principal chiral model, the $S$-matrices are
constructed from group-invariant solutions of the YBE. The spectrum of the
theory then consists of multiplets within which the particles have equal mass
and which form representations of the group $G$.

One method for constructing these $S$-matrices is the bootstrap procedure
(known as the `fusion procedure' for solutions of the YBE) in which, at
appropriate
poles, intermediate states of the $S$-matrix are identified as particle states
whose $S$-matrices can then be constructed. An alternative method is to
construct explicitly the action of the underlying charge algebra on particle
multiplets, and then use conservation of these charges to deduce the
$S$-matrix.
Bernard\cite{Bern} recently showed that this algebra is precisely
Drinfeld's
Yangian\cite{D} $Y(\A)$, where $\A$ is the Lie algebra of the group $G$:
if we
write the action of the charges on states consisting of two asymptotically
free
particles as $\Delta$, there is a local charge satisfying
\begin{equation}\label{cr0}
\left[ Q_0^a , Q_0^b \right] = i \hbar f^{abc} Q_0^c
\hspace{0.3in}{\rm and}\hspace{0.3in}
\Delta(Q_0^a) = Q_0^a \otimes 1 + 1 \otimes Q_0^a \hsp,
\end{equation}
where $f^{abc}$ are the structure constants of $\A$,
and a series of non-local charges, the first of which satisfies
\begin{equation}\label{cr1}
\left[ Q_0^a , Q_1^b \right] = i \hbar f^{abc} Q_1^c  \hspace{0.3in}{\rm
and}\hspace{0.3in}
\Delta(Q_1^a) = Q_1^a \otimes 1 + 1 \otimes Q_1^a + {1\over  2}f^{abc} Q_0^c
\otimes Q_0^b \hsp.
\end{equation}
The requirement that $\Delta$ be a homomorphism ({\em i.e.\ }that asymptotic
states carry representations of the charge algebra) fixes\footnote{For $\A\neq
sl(2)$. For the general condition see Drinfeld\cite{D}.}
\begin{equation}\label{YSerre}
 f^{d[ab} [Q_1^{c]},Q_1^d] \hsp = \hsp {i\hbar\over {12}} \,f^{ap i} f^{bq
j}f^{cr k}f^{ijk} \, Q_0^{(p} Q_0^q Q_0^{r)} \hsp,
\end{equation}
where $[\,]$ and $(\,)$ denote (anti-)symmetrization on the enclosed indices.

This charge algebra provides a dynamical symmetry since it does not commute
with the Poincar\'e group: if the rapidity (defined by ${\bf
p}=(m\,{\rm{cosh}}\theta,m{\rm\,{sinh}}\theta)$) of a state is given a Lorentz
boost $L_\theta$ of rapidity $\theta$, it is found that
\begin{equation}\label{Lorentz}
Q_0^a\mapsto Q_0^a \hspace{0.3in}{\rm and}\hspace{0.3in}Q_1^a \mapsto Q_1^a -
{{\hbar C_{Adj}}\over {4\pi}} \, \theta \, Q_0^a
\end{equation}
where $C_{Adj}\delta_{ad} = f^{abc}f^{cbd}$ gives the value of the quadratic
Casimir operator $C_2=Q_0^a Q_0^a$ in the adjoint representation.
The $S$-matrix is thus constrained by conservation of $Q_0^a$ and $Q_1^a$, so
that
the interaction  $S(\theta_1-\theta_2)$ of two particles of rapidities
$\theta_1$ and $\theta_2$ satisfies
\begin{eqnarray}\label{sm1}
 & \left[ \,S(\theta_1-\theta_2) \, , \, Q_0^a \otimes 1 + 1 \otimes Q_0^a
\,\right] = 0 & \\[0.1in] \label{sm2}
 & S(\theta_1-\theta_2) \left( L_{\theta_1} \otimes L_{\theta_2} \Delta(Q_1^a)
\right) = \left( L_{\theta_2} \otimes L_{\theta_1} \Delta(Q_1^a) \right)
S(\theta_1-\theta_2) \hsp, &
\end{eqnarray}
which together imply the Yang-Baxter equation
$$
(S(\theta_2-\theta_3)\otimes 1)\,(1\otimes
S(\theta_1-\theta_3))\,(S(\theta_1-\theta_2)\otimes 1) = (1\otimes
S(\theta_1-\theta_2))\,(S(\theta_1-\theta_3)\otimes 1)\,(1\otimes
S(\theta_2-\theta_3) \hsp.
$$
This acts on a state consisting of three asymptotically free particles of
rapidities $\theta_1$, $\theta_2$ and $\theta_3$, with each factor giving an
interaction between two of the three particles. The $S$-matrix is related to
the usual Yangian $R$-matrix by $S={\bf P}R$, where ${\bf P}$ transposes states
in a tensor product.

The particle multiplets of the theory are then irreducible representations of
$Y(\A)$ which may or may not be irreducible as representations of the Lie
subalgebra $\A$, (\ref{cr0}). In particular, an irreducible representation $V$
of $\A$ may be extended to a representation
$v$ of $Y(\A)$, with
\begin{equation}\label{rep}
\rho_v(Q_0)=\rho_V(Q_0)  \hspace{0.3in} {\rm and} \hspace{0.3in} \rho_v(Q_1)=0
\hsp,
\end{equation}
provided the action of the right-hand side of (\ref{YSerre}) vanishes on $V$,
and Drinfeld classified those $V$ for which this is true\cite{D}. In
particular, it is true of all the fundamental representations of $a_n$ and
$c_n$, but of only a few of the fundamental representations of other algebras.
Since we expect the particle multiplets to be fundamental representations of
$Y(\A)$
({\em i.e.\ }those representations of $Y(\A)$ containing a
fundamental
representation of $\A$ as a component), we therefore have an explicit
action of
the charge algebra on all the particle multiplets of the $a_n$ and
$c_n$
theories: the charges have action (\ref{rep}) on particles of zero
rapidity,
boosted by (\ref{Lorentz}) on a particle of rapidity $\theta$.

A method then exists\cite{ratr} for solving (\ref{sm1},\,\ref{sm2}). For full
details we refer the reader to the original papers, but in brief one first uses
the group invariance (\ref{sm1}) of $S(\theta)$ and Schur's lemma to give, for
irreducible $\A$-representations $X$ and $Y$ where $X\otimes Y$ contains no
multiplicities,
$$
S_{XY}(\theta) = {\bf P} \sum_{W \subset X \otimes Y} \tau_W(\theta) P_W \hsp,
$$
where the $P_W$ project onto irreducible components of $X\otimes Y$. We next
use
$$
{1\over  2} [ C_2, 1\otimes Q_0^a ] = f^{abc} Q_0^c \otimes Q_0^b
$$
and project on the left and right of (\ref{sm2}) with $P_R$ and $P_S$ to obtain
\begin{equation}\label{tau}
  {{\tau_{S}(u)}\over {\tau_{R}(u)}} = {{\theta+ \Delta_{RS}}\over
{\theta-\Delta_{RS}}}
\end{equation}
where
$$
\Delta_{RS} = {C_2(R)-C_2(S)\over C_{Adj}} \,i\pi
$$
for those $R,S$ which have opposite parity in $X\otimes Y$ and for which $R
\subset adjoint \otimes S$. This system of equations is made most transparent
by letting the components of $X\otimes Y$ be the nodes of a bipartite graph,
joined by directed edges $R\rightarrow S$ labelled with the differences of the
Casimir operators when there is a corresponding equation (\ref{tau}). That the
system is consistent is then equivalent to the requirement that the set of
labels on the
edges be the same for all paths between two nodes of the graph. As expected
from
Drinfeld's results, this is found to be the case for all the fundamental
representations
of $a_n$ and $c_n$. The solutions for the $a_n$ case were obtained by Kulish,
Sklyanin and Reshetikhin\cite{KSR} using both the fusion procedure and a method
similar to this, but for $c_n$ only a few solutions have been found:
with the fundamental representations of $c_n$ labelled by

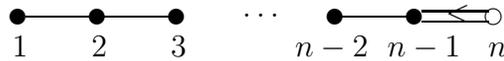
\vspace{0.2in}
\begin{picture}(250,20)(-90,0)
\put(20,15){\circle*{6}}
\put(50,15){\circle*{6}}
\put(80,15){\circle*{6}}
\put(140,15){\circle*{6}}
\put(170,15){\circle*{6}}
\put(200,15){\circle{6}}
\put(23,15){\line(1,0){24}}
\put(53,15){\line(1,0){24}}
\put(143,15){\line(1,0){24}}
\put(173,13){\line(1,0){25}}
\put(173,17){\line(1,0){25}}
\put(182,13){$<$}
\put(105,15){$\dots$}
\put(18,0){$1$}
\put(48,0){$2$}
\put(78,0){$3$}
\put(125,0){$n-2$}
\put(160,0){$n-1$}
\put(198,0){$n$}
\end{picture}

\parindent 0pt
the bootstrap procedure has been used\cite{ORW} to construct $S_{1m}(\theta)$
starting from the $S_{11}(\theta)$ previously calculated\cite{BKWK}. However,
we can now construct the full set $S_{lm}(\theta)$ using the tensor product
graph (TPG), which for $l\otimes m$, $l\geq m$ is
\pagebreak
$$
\begin{array}{ccccccccccc}
\lambda_l+\lambda_m & \rightarrow & \lambda_{l+1}+\lambda_{m-1}& \cdots &
\rightarrow & \lambda_n+\lambda_{l+m-n} &  \cdots & \rightarrow &
\lambda_{l+m-1}+\lambda_1 & \rightarrow & \lambda_{l+m}\\[0.1in]
\downarrow & & \downarrow & & & \downarrow & & & \downarrow \\[0.1in]
\lambda_{l-1}+\lambda_{m-1} & \rightarrow & \lambda_l+\lambda_{m-2} & \cdots &
\rightarrow & \lambda_{n-1}+\lambda_{l+m-n-1} &  \cdots & \rightarrow &
\lambda_{l+m-2}\\[0.1in]
\vdots & & \vdots & & & \vdots \\[0.05in]
\downarrow & & \downarrow & & & \downarrow \\[0.1in]
\lambda_{n-m} + \lambda_{n-l}& \rightarrow & \lambda_{n-m+1}+\lambda_{n-l-1}&
\cdots & \rightarrow & \lambda_{2n-l-m} \\[0.1in]
\vdots & & \vdots \\[0.05in]
\downarrow & & \downarrow \\[0.1in]
\lambda_{l-m+1}+\lambda_1 & \rightarrow & \lambda_{l-m+2} \\[0.1in]
\downarrow \\[0.1in]
\lambda_{l-m} \\[0.1in]
\end{array}
$$
where the representations are indicated by their highest weights, given in
terms of the fundamental weights $\lambda_i$, and $\lambda_0 \equiv 0$
indicates a singlet. (In addition, to keep the graph simple, the labels
have been left out.) For $l+m>n$ the graph truncates at the $(n-l+1)$th
column, since the representations to the right of this column in the
graph are then no longer present in the decomposition of $l\otimes m$.

\parindent 12pt
The $S$-matrix is then given by
\begin{equation}\label{matrix}
S_{lm}(\theta) = {\bf P}\, s_{lm}(\theta) \sum_{p=0}^{{\rm Min}(n-l,m)}
\sum_{q=0}^{m-p} \tau_{\lambda_{l+p-q}+\lambda_{m-p-q}}(\theta)
P_{\lambda_{l+p-q}+\lambda_{m-p-q}}
\end{equation}
where
\begin{equation}\label{tau2}
\tau_{\lambda_{l+p-q}+\lambda_{m-p-q}}(\theta) = \prod_{p'=1}^p [2p'+l-m]
\prod_{q'=1}^q [h+2q'-l-m]
\end{equation}
with
$$
[x] \equiv {{\theta + {x i \pi\over h}}\over{\theta - {x i \pi\over h}}}
\hspace{0.4in}{\rm and}\hspace{0.4in} h=2n+2 \hsp.
$$
In (\ref{matrix}), $s_{lm}(\theta)$ is an overall scalar factor which cannot be
determined from the TPG. However, we can fix $s_{11}(\theta)$ up to an overall
CDD ambiguity by requiring that $S_{11}(\theta)$ be unitary and
crossing-symmetric. A solution that achieves this in such a way that
$S_{11}(\theta)$ has no poles in the physical strip $0\leq {\rm Im}\,\theta
\leq \pi$ is\cite{ORW}

$$
s_{11}(\theta) = k(\theta) = {{ \Gamma({i\theta\over2\pi})
\Gamma(-{i\theta\over2\pi}+{1\over2}) \Gamma(-{i\theta\over2\pi}+ {1\over h})
\Gamma({i\theta\over2\pi} +{1\over h} +{1\over2})
}\over{\Gamma(-{i\theta\over2\pi}) \Gamma({i\theta\over2\pi}+{1\over2})
\Gamma({i\theta\over2\pi}+ {1\over h}) \Gamma(-{i\theta\over2\pi} +{1\over h}
+{1\over2}) }} \hsp.
$$

Instead, however, we assume some bound state structure and choose a
solution\cite{ORW} which has a positive residue (direct channel) simple pole
at $\theta={2i\pi\over h}$ corresponding to the particle fusion
$11\rightarrow 2$, which is possible because at this value of $\theta$
the $S$-matrix projects only on to the $2$ component of $1\otimes 1$ -
as can easily be seen from the TPG for $1\otimes 1$,
$$
\begin{array}{ccc}
2\lambda_1 & \longrightarrow^{2i\pi\over h} & \lambda_2 \\[0.1in]
\downarrow_{i\pi} \\[0.1in]
\lambda_0 \\
\end{array}
$$
A solution which achieves this is then
$$
s_{11}(\theta) = - (2) (h-2) k(\theta)
$$
where we have used the notation\cite{BCDS1}
$$
(x) \equiv {{{\rm{sinh}}({\theta\over2}+{{i\pi
x}\over{2h}})}\over{{\rm{sinh}}({{\theta}\over2}-{{i\pi x}\over{2h}})}} \hsp.
$$
In principle, we could now have deduced all the $S_{lm}(\theta)$ from
$S_{11}(\theta)$ using the bootstrap principle, which implies that
$$
S_{tr}(\theta)= \left. \left( 1\otimes S_{tm}(\theta + i
\bar{\theta}^{l}_{rm})\right)\left(S_{tl}(\theta -
i\bar{\theta}^{m}_{rl})\otimes 1 \right) \right|_{r} \hsp,
$$
where $\bar{\theta}=\pi-\theta$ and $|_r$ indicates the restriction of the
tensor product $l\otimes m$ to state $r$, or, schematically,

\hspace{10mm}
\setlength{\unitlength}{1mm}
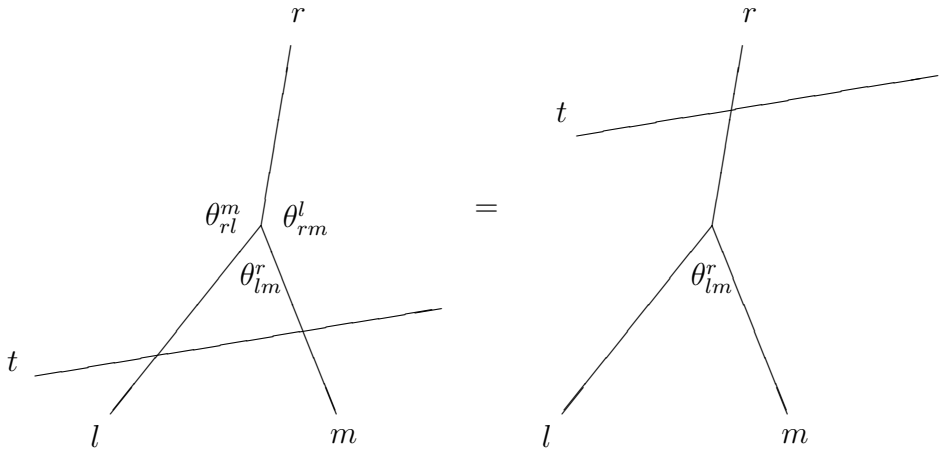
\begin{picture}(130,65)(-65,-35)
\put(-30,-5){\line(-4,-5){20}}
\put(-30,-5){\line(2,-5){10}}
\put(-30,-5){\line(1,6){4}}
\put(30,-5){\line(-4,-5){20}}
\put(30,-5){\line(2,-5){10}}
\put(30,-5){\line(1,6){4}}
\put(-60,-25){\line(6,1){54}}
\put(12,7){\line(6,1){48}}
\put(0,-3){\makebox(0,0){=}}
\put(-52,-33){\makebox(0,0){$l$}}
\put(-19,-33){\makebox(0,0){$m$}}
\put(8,-33){\makebox(0,0){$l$}}
\put(41,-33){\makebox(0,0){$m$}}
\put(35,23){\makebox(0,0){$r$}}
\put(-25,23){\makebox(0,0){$r$}}
\put(-63,-23){\makebox(0,0){$t$}}
\put(10,10){\makebox(0,0){$t$}}
\put(-30,-12){\makebox(0,0){${\theta}^{r}_{lm}$}}
\put(30,-12){\makebox(0,0){${\theta}^{r}_{lm}$}}
\put(-35,-4){\makebox(0,0){${\theta}^{m}_{rl}$}}
\put(-24,-4){\makebox(0,0){${\theta}^{l}_{rm}$}}
%\put(30,7){\makebox(0,0){${\theta}$}}

\end{picture}

\parindent 0pt
whenever $S_{lm}(\theta)$ has a positive residue simple pole at
$i{\theta}^{r}_{lm}$ with residue proportional to $r\subset l\otimes m$.
However, in practice the calculations, which involve complex computations in
Brauer's algebra (the centralizer algebra of $c_n$) are too complicated;
we must instead use the TPG to give the matrix structure. However, we {\em can}
use the bootstrap to determine the scalar factors $s_{lm}(\theta)$. We
then have
\begin{equation}\label{s}
s_{lm}(\theta) = X_{lm}(\theta) k_{lm}(\theta)
\end{equation}
where
\begin{equation}\label{k}
k_{lm}(\theta) = \prod_{\stackrel{p=1-m}{step\,2}}^{m-1}
\prod_{\stackrel{q=1-n}{ step\,2}}^{n-1} k(\theta + {p+q\over h}i\pi)
\end{equation}
and the $X_{lm}$, which give all the pole structure of the $S$-matrices, have
in fact already been calculated in the context of purely elastic
scattering theories (in which the particles do not form multiplets), where
they are
related to $d_{n+1}^{(2)}$ affine Toda theories\cite{BCDS1}:
\begin{equation}\label{X}
X_{lm}(\theta) = \prod_{\stackrel{p=l-m+1}{step\,2}}^{l+m-1}
(p-1)(p+1)(h-p-1)(h-p+1) \hsp.
\end{equation}
Thus the complete set of $S_{lm}(\theta)$ is given by substituting (\ref{tau2})
and (\ref{s},\,\ref{k},\,\ref{X}) into (\ref{matrix}).

\parindent 12pt
Since the matrix structure of $S_{lm}$ has been found by a method other
than the bootstrap, it is worth examining how the bootstrap would work
for our solutions. As in the $S_{11}$ case, we can look at the TPG
for $l \otimes m$ and, seeing that the representations $l-m$ and (for
$l+m \leq n$) $l+m$ are connected to the rest of the graph by a single edge
whose label is valued in the physical strip, we expect $S_{lm}$ to
have bootstrap poles corresponding to these fusings. This is indeed
what we find: such fusings correspond precisely to the positive
residue simple poles in $X_{lm}$, and the bootstrap procedure on
these poles thus closes on the expected spectrum of $n$ massive
multiplets. This correspondence is really quite remarkable -
facts about solutions of the YBE are being predicted by the bootstrap
structure of {\em scalar} functions $X_{lm}$. We can think of the
fusings as being due to a three-point coupling between the particles
$l$, $m$ and $l+m$ with
\begin{equation}\label{angles}
\theta^{l+m}_{l\,m}= {l+m\over h}\pi \hsp, \hspace{0.25in}
\theta^{l}_{l+m\,m}={h-l\over h}\pi \hspace{0.2in} {\rm and}
\hspace{0.2in}\theta^{m}_{l+m\,l}={h-m\over h}\pi \hsp,
\end{equation}
and then the fact that $\theta^{l+m}_{l\,m} + \theta^{l}_{l+m\,m}
+\theta^{m}_{l+m\,l}= 2 \pi$ is a highly non-trivial consequence of Yangian
representation theory, proved
only in the very special case of a three point coupling between identical
particles\cite{CP3}.

There is, however, a subtlety in that there are also positive residue cubic
poles
in $X_{lm}(\theta)$ for which the residue of $S_{lm}(\theta)$ does
not correspond to a fundamental representation or indeed to any subgraph
of the TPG. We therefore expect that such poles should not be interpreted
as bootstrap poles, yet we know from affine Toda theories\cite{BCDS1}
and the $d_4$ Yangian-invariant theory\cite{facsm} that such poles
{\em can} mask simple poles in particle states and thus form part of the
bootstrap. Thus, at present, we are forced to fall back on the
{\em postulate} that the spectrum of the theory consists only
of particles in fundamental representations of the Yangian.
It would therefore be nice to have an independent, quantum field theoretic
way of deducing whether or not a given non-simple pole in $S_{lm}(\theta)$
should
be included in the bootstrap. Whereas in affine Toda theories direct
comparison with perturbation theory is possible\cite{BCDS1}, for models
with multiplet structure (such as the principal chiral model) matters
are more complicated, and methods such as the ${1\over N}$-expansion
do not seem promising.

The spectrum of masses in the theory can be deduced from conservation of
momentum at bootstrap poles, and for the $c_n$ theories has been
deduced\cite{ORW} from
$S_{1m}(\theta)$:
\begin{equation}\label{mass}
m_p = m \sin \left( {p \pi\over h} \right) \hsp, \hspace{0.4in}
p=1,\dots,n\hsp.
\end{equation}
An interesting alternative way of deducing the values of the bootstrap poles
and thus the mass spectrum has been proposed recently by Belavin\cite{Bel} and
used by him to compute the $a_n$ mass spectrum.
He considered the two commuting conserved charges $Q_0^a Q_0^a\,(=C_2)$ and
$Q_1^a Q_0^a$
and applied the bootstrap principle: if the residue of a pole of
$S_{lm}(\theta)$ is a particle state $r$ then
\begin{eqnarray*}
\Delta(Q_0^a)\Delta(Q_0^a) \, l\otimes m & = & Q_0^a Q_0^a \, l\otimes m |_r \\
\Delta(Q_1^a)\Delta(Q_0^a) \, l\otimes m & = & Q_1^a Q_0^a \, l\otimes m |_r
\hsp.
\end{eqnarray*}
Using the coproducts (\ref{cr0},\,\ref{cr1}) and the representation
(\ref{rep}),
and setting $\theta_r=0$, it is then possible to deduce the
value of $i\theta_{lm}^{r} = \theta_{l}-\theta_{m}$:
$$
\theta_{lm}^{r} = { {2\tilde{r}\,( \tilde{l} + \tilde{m} - \tilde{r} )} \over {
2( \tilde{l}\tilde{r} + \tilde{m} \tilde{r} + \tilde{l} \tilde{m} ) -
(\tilde{l}^2 + \tilde{m}^2 + \tilde{r}^2 ) }} \pi \hsp,
$$
where we have written $\tilde{p} \equiv C_2(p)$. When we apply this method
to the $c_n$ case we obtain the expected fusing angles (\ref{angles}) and
thus (\ref{mass}); Belavin's method also works in this way for all
particle multiplets (for any $\A$) which are irreducible as representations of
$\A$.

If we wish to develop the methods discussed in this paper as an alternative
to the bootstrap procedure for calculating factorized $S$-matrices for
general $\A$, it is clear that new results on representations of Yangians
are needed\footnote{At this point we should note that recent results\cite{SL}
on off-shell, infinite-dimensional representations of dynamical Yangian
symmetry do not seem to help with this.}: both the TPG and Belavin's
method depend crucially on the explicit action (\ref{rep}) of $Q_0^a$
and $Q_1^a$ on the particles. Apart from $v=V$, the only case for which
such an action is known is Drinfeld's construction\cite{D} of $v=adjoint
\oplus singlet$; the corresponding $R$-matrices have been constructed
by Chari and Pressley\cite{CP3}, although it is not clear how to
extend Belavin's method to this representation.

Finally, it seems that neither the bootstrap (which describes the
decomposition of tensor products of $Y(\A)$-representations)
nor
the methods\cite{ratr,CP3} for solving (\ref{sm2}) give any
general
insight into the mass spectra
and fusings obtained. Since, as we
mentioned above, much information
about the YBE is already contained in
the $X_{lm}$, and since the mass
spectra and fusings given by the $X_{lm}$
have a beautiful description in
terms of root systems of Lie
algebras\cite{ped} (at least for
simply-laced $\A$; for
non-simply-laced $\A$ the situation
is more complicated), it
therefore appears that there is
every prospect of rich undiscovered
structure in the representation
theory of the Yangian.

\vfill
\pagebreak

\end{document}